\documentclass[10pt, conference, compsocconf]{IEEEtran}
\IEEEoverridecommandlockouts

\def\BibTeX{{\rm B\kern-.05em{\sc i\kern-.025em b}\kern-.08em
T\kern-.1667em\lower.7ex\hbox{E}\kern-.125emX}}

\usepackage{algorithmic}
\usepackage{amsmath,amssymb,amsfonts}
\usepackage{authblk}
\usepackage{balance}
\usepackage{booktabs}
\usepackage[dvipsnames]{xcolor}
\usepackage{colortbl}
\usepackage[T1]{fontenc}
\usepackage{graphicx}
\usepackage[colorlinks,
    linkcolor=blue,
    citecolor=blue,
    pdftex,
    pdfauthor={assert},
    pdftitle={proze},
    pdfsubject={generation of real tests with real test data and incredible oracles},
    pdfkeywords={assertions, puts, triskaidekaphobia},
    pdfcreator={gru}]{hyperref}
\usepackage[utf8]{inputenc}
\usepackage{listings}
\usepackage{makecell}
\usepackage{multirow}
\usepackage{soul}
\usepackage[most]{tcolorbox}
\usepackage{textcomp}
\usepackage{xspace}
\usepackage{biblatex}
\AtEveryBibitem{\clearfield{pages}} 

\bibliography{main}

\newcommand{\proze}{\textsc{proze}\xspace}

\newcommand{\etal}{\textit{et al.}\xspace}

\newcommand{\pdfboxproject}{\textsc{PDFBox}\xspace}

\newcommand{\saml}{\textsc{wso2-saml}\xspace}

\newcommand{\fontbox}{\texttt{fontbox}\xspace}
\newcommand{\xmpbox}{\texttt{xmpbox}\xspace}
\newcommand{\pdfbox}{\texttt{pdfbox}\xspace}
\newcommand{\query}{\texttt{query}\xspace}
\newcommand{\sso}{\texttt{sso}\xspace}

\newcommand{\greyassertion}{\badge[lightgray]{strongly-coupled}\xspace}
\newcommand{\goldassertion}{\badge[yellow]{falsifiably-coupled}\xspace}
\newcommand{\greenassertion}{\badge[Aquamarine]{decoupled}\xspace}

\newcommand{\nbmodules}{5\xspace}
\newcommand{\nbtargets}{128\xspace}

\newcommand{\nbputs}{2,287\xspace}
\newcommand{\nbcuts}{248\xspace}
\newcommand{\nbgreys}{686\xspace}
\newcommand{\nbgreens}{1,011\xspace}
\newcommand{\nbreds}{373\xspace}
\newcommand{\nbgolds}{217\xspace}

\newcommand{\nbpdfputs}{2,170\xspace}
\newcommand{\nbpdfcuts}{206\xspace}
\newcommand{\nbpdfTargetMethods}{94\xspace}

\newcommand{\nbssogreys}{51\xspace}
\newcommand{\nbssogreens}{24\xspace}
\newcommand{\nbssogolds}{4\xspace}
\newcommand{\nbssoputs}{109\xspace}
\newcommand{\nbssocuts}{39\xspace}
\newcommand{\nbssotargetMethods}{32\xspace}

\newcommand{\nbquerygreys}{0\xspace}
\newcommand{\nbquerygreens}{7\xspace}
\newcommand{\nbquerygolds}{0\xspace}
\newcommand{\nbqueryputs}{8\xspace}
\newcommand{\nbquerycuts}{3\xspace}
\newcommand{\nbquerytargetMethods}{2\xspace}

\definecolor{rubyred}{RGB}{169,20,1}
\definecolor{fakerred}{RGB}{222,63,36}
\definecolor{commentcolor}{HTML}{626262}
\definecolor{pteal}{RGB}{0,128,128}
\definecolor{stringcolor}{HTML}{17c6a3}
\definecolor{keywordcolor}{HTML}{dd2867}
\definecolor{classcolor}{HTML}{1290c3}
\definecolor{numbercolor}{HTML}{6897bb}
\definecolor{dim}{rgb}{0.55, 0.57, 0.67}
\definecolor{light-gray}{gray}{0.95}
\definecolor{pgreen}{RGB}{5,205,107}
\definecolor{pblue}{RGB}{2,154,223}

\lstdefinestyle{java}{
    language={java},
    basicstyle=\ttfamily\footnotesize, 
    morekeywords={String, @Test, @ParameterizedTest, @MethodSource, @CsvSource, assertEquals, assertTrue},
    keywordstyle=\bfseries\color{rubyred},
    emph={},
    emphstyle={\bfseries\itshape\color{fakerred}},
    stringstyle=\bfseries\color{teal},
    commentstyle=\itshape\color{fakerred},
    breaklines=true,
    showlines=true,
    captionpos=b,
    float=tb,
}
\newcommand{\hlgreen}[1]{%
    {%
    \sethlcolor{green!30}%
    \hl{#1}%
    }%
}

\newtcbox{\badge}[1][red]{
  on line, 
  arc=2pt,
  colback=#1,
  colframe=#1!50!white,
  fontupper=\color{black},
  boxrule=1pt, 
  boxsep=0pt,
  left=1pt,
  right=1pt,
  top=1pt,
  bottom=1pt
}

\usepackage[framemethod=tikz]{mdframed}
\mdfdefinestyle{mpdframe}{
    frametitlebackgroundcolor   =cyan!20,
    frametitlefont              =\itshape,
    frametitlerule              =true,
    roundcorner                 =1pt,
    middlelinewidth             =1pt,
    innermargin                 =0.1cm,
    outermargin                 =0.1cm,
    innerleftmargin             =0.1cm,
    innerrightmargin            =0.1cm,
    innertopmargin              =0.1cm,
    innerbottommargin           =0.1cm,
    linecolor                   =cyan,
    skipabove                   =6pt
}

\newcommand{\revised}{\textcolor{black}}

\begin{document}
\title{\proze: Generating Parameterized Unit Tests \\ Informed by Runtime Data}

\author[*]{Deepika Tiwari}
\author[$\dag$]{Yogya Gamage}
\author[*]{Martin Monperrus}
\author[$\dag$]{Benoit Baudry}
\affil[*]{KTH Royal Institute of Technology, \textit{\{deepikat, monperrus\}@kth.se}}
\affil[$\dag$]{Université de Montréal, \textit{\{yogya.gamage, benoit.baudry\}@umontreal.ca}}

\maketitle

\begin{abstract}
Typically, a conventional unit test (CUT) verifies the expected behavior of the unit under test through one specific input / output pair.
In contrast, a parameterized unit test (PUT) receives a set of inputs as arguments, and contains assertions that are expected to hold true for all these inputs.
PUTs increase test quality, as they assess correctness on a broad scope of inputs and behaviors.
However, defining assertions over a set of inputs is a hard task for developers, which limits the adoption of PUTs in practice.
In this paper, we address the problem of finding oracles for PUTs that hold over multiple inputs.

We design a system called \proze, that generates PUTs by identifying developer-written assertions that are valid for more than one test input.
We implement our approach as a two-step methodology: first, at runtime, we collect inputs for a target method that is invoked within a CUT; next, we isolate the valid assertions of the CUT to be used within a PUT.

We evaluate our approach against \nbmodules real-world Java modules, and collect valid inputs for \nbtargets target methods, from test and field executions. 
We generate \nbputs PUTs, which invoke the target methods with a significantly larger number of test inputs than the original CUTs. 
We execute the PUTs and find \nbgolds that provably demonstrate that their oracles hold for a larger range of inputs than envisioned by the developers.
From a testing theory perspective, our results show that developers express assertions within CUTs, which actually hold beyond one particular input.
\end{abstract}


\section{Introduction}
Within all serious software projects, developers write hundreds of unit tests that automatically verify the expected behaviors of individual components in their systems \cite{karhu2009empirical}.
As part of their testing effort, developers spend time deciding on concrete pairs of test input and oracle \cite{barr2014oracle}.
The input brings the system to the desired testable state through a series of actions, such as method invocations.
The oracle verifies the expected behavior, and is typically expressed through assertion statements.
Conventional unit tests (CUTs) assess the behavior of the program with respect to one specific input - output pair. 

Parameterized unit tests (PUTs) are a more advanced form of tests.
Unlike CUTs, PUTs evaluate multiple test inputs against the oracle \cite{tillmann2005parameterized}, exercising more behaviors of the unit under test \cite{fraser2011generating, soares2022refactoring}.
Writing PUTs is notoriously hard for developers, which is why they are not widely used in practice \cite{thummalapenta2011retrofitting, tsukamoto2018autoput}.
The major blockers for writing PUTs are envisioning the appropriate oracle and defining representative inputs \cite{goldstein2024property}.
In this paper, our key insights are that certain assertions already present in conventional unit tests are suitable for parameterized unit tests \cite{saff2008theory}, and that runtime data can be utilized within parameterized unit tests \cite{wang2017behavioral}.

We present \proze, a novel system that analyzes CUTs in order to generate PUTs.
\proze automatically detects assertions that are suitable for PUTs, according to the following methodology.
First, for a target method that is directly invoked by a CUT, it captures additional input arguments from its invocations at runtime, across test and field executions of the program.
Next, \proze transforms the candidate CUT into PUTs that invoke the target method with its arguments captured at runtime.
Each generated PUT contains a developer written assertion that is evaluated against all captured inputs.
We categorize each PUT output by \proze as:
1) \greyassertion if it is valid only for the original test input and cannot generalize to other inputs; 
2) \greenassertion if it is valid for all observed input values;
and 3) \goldassertion if it is valid for a subset of input values and fails for others.

We evaluate \proze against \nbmodules modules from real-world Java projects.
\proze monitors the execution of the developer-written test suite of each module, as well as its execution under a usage workload, to capture new test inputs, and synthesize parameterized unit tests.
From our experiments with the \nbmodules modules, we observe that \proze successfully generates \nbputs PUTs which compile and can be run with off-the-shelf test harnesses. 
We find \nbgolds PUTs that are provably valid over a broader range of inputs.

\proze is fundamentally novel.
We are aware of only a few papers exploring the field of PUT generation: Fraser \etal concentrate on symbolic pre- and postconditions \cite{fraser2011generating}, Thummalapenta \etal look at simple tests \cite{thummalapenta2011retrofitting}, and Elbaum \etal. consider system tests  \cite{elbaum2009carving}.
Our work is the first to generalize CUTs into PUTs for real world, complex tests, using runtime data.
Our key contributions are:
\begin{itemize}
    \item \proze, a novel methodology to 
    generate parameterized unit tests by identifying developer-written assertions that hold over multiple inputs collected at runtime.
    \item Evidence that 1) some real-world assertions hold for multiple inputs beyond the original ones, and 2) \proze is able to generate executable parameterized unit tests that increase input space coverage by several orders of magnitude.
    \item A publicly available implementation of the approach, which is able to generate parameterized unit tests for modern Java software and testing frameworks.
\end{itemize}

\section{Background}\label{sec:background}
We now introduce and illustrate the two main concepts for our work: conventional unit tests and parameterized unit tests.
We reuse the terminology of conventional vs. parameterized introduced in the seminal series of works by de Halleux, Tillman, and Xie \cite{xie2009mutation}.

\begin{lstlisting}[style=java, showlines=true, label={lst:bg-cut-put}, caption={\texttt{testOneElement} is a conventional unit test (CUT) with three assertions evaluated against a single input.
\texttt{testManyElements} is a parameterized unit test (PUT) which receives a \texttt{String} argument from the argument provider \texttt{provideSymbol}.
Its three assertions hold over the 7 unique inputs supplied by \texttt{provideSymbol}.
}, float]
// CUT with an oracle coupled to the input
@Test
public void testOneElement() {
  PeriodicTable table = PeriodicTable.create();
  Element element = table.getElement("He");
  assertEquals("Helium", element.getFullName());
  assertEquals(4, element.calculateMassNumber());
  assertTrue(table.getGroup18().contains(element));
}%*\par\noindent\dotfill*)

// PUT with a more general oracle
@ParameterizedTest
@MethodSource("provideSymbol")
public void testManyElements(String symbol) {
  PeriodicTable table = PeriodicTable.create();
  Element element = table.getElement(symbol);
  int atomicNum = element.getAtomicNumber();
  int massNum = element.calculateMassNumber();
  int numNeutrons = element.getNeutrons();
  assertEquals(numNeutrons, (massNum - atomicNum));
  assertTrue(atomicNum < massNum);
  assertTrue(table.getGroup18().contains(%*\href{https://en.wikipedia.org/wiki/List_of_fictional_elements,_materials,_isotopes_and_subatomic_particles}{element}*)));
}

// Argument provider for the PUT
private static Stream<Arguments> provideSymbol() {
  return Stream.of(
    Arguments.of("He"),
    Arguments.of("Ne"),
    Arguments.of("Ar"),
    Arguments.of(%*\href{https://www.supermanhomepage.com/tv/tv.php?topic=articles/seinfeld}{"Kr"}*)),
    Arguments.of("Xe"),
    Arguments.of("Rn"),
    Arguments.of("Og")
  );
}
\end{lstlisting}

\subsection{Conventional Unit Tests (CUT)}
A CUT is a developer-written test case that typically invokes the unit under test with a single input state, and verifies the expected output state through an oracle \cite{xie2009mutation}.
One single CUT can test the behavior of multiple methods that belong to the unit under test, and can define multiple assertions to check the behavior of these methods for a specific test input \cite{white2022tctracer}. 

Lines 1 to 9 of \autoref{lst:bg-cut-put} present the CUT \texttt{testOneElement}, annotated with the \texttt{@Test} JUnit annotation.
The method call \texttt{getElement} within the CUT (line 5) fetches the element with the symbol \texttt{"He"}, from an instance of \texttt{PeriodicTable}.
The CUT contains three assertions, verifying the expected properties related to the element.
The first and second assertions verify its full name and its mass number, respectively.
These two assertions are specific to the given input, and will not hold if \texttt{getElement} is invoked with any other \texttt{String} argument.
The third assertion verifies that the element is contained within the set of elements in \href{https://en.wikipedia.org/wiki/Noble_gas}{Group 18} of the \texttt{table}. 
This is a more general assertion, which will hold for all Group 18 elements.

\subsection{Parameterized Unit Tests (PUT)}
A PUT is a test case that receives one or more parameters \cite{kampmann2019carving}, and that invokes the unit under test with multiple inputs corresponding to the input parameters.
A PUT typically expresses a more general oracle than a CUT, by design, since it must hold over a set of behaviors of the unit under test.
The parameters of a PUT are automatically assigned values through an \emph{argument provider}.
The oracle is evaluated each time the PUT is invoked with a new argument.

\emph{Argument provider}: An argument provider can be an array of values or a helper method that returns values.
The argument provider for a PUT returns values that are of the same type as the type of the PUT's parameters.
The PUT is invoked as many times as there are values in the argument provider. 

Consider the method \texttt{testManyElements} of \autoref{lst:bg-cut-put} (lines 12 to 24).
The JUnit annotation \texttt{@ParameterizedTest} declares that this method is a PUT \cite{soares2022refactoring}.
Next, the annotation \texttt{@MethodSource} links it to the argument provider method \texttt{provideSymbol} (lines 26 to 37). 
When the PUT is run, its \texttt{String} parameter \texttt{symbol} is assigned one value supplied by \texttt{provideSymbol}, resulting in 7 unique runs of the PUT in total.
Within the PUT, the method \texttt{getElement} gets invoked with the \texttt{symbol} (line 17), and returns the corresponding element from the \texttt{table}.
None of the three assertions in the PUT are specific to the input.
The first and second assertions verify  properties related to the atomic and mass numbers of the element, computed within the test.
The third assertion is the same as the third assertion of \texttt{testOneElement}, and checks that the element is contained within Group 18 of the \texttt{table}. 
All three assertions hold over all the 7 \texttt{String} input arguments supplied by the argument provider. 

In this work, we hypothesize that some developer-written CUTs include an oracle that is meant to be valid over more than one test input \cite{saff2008theory}.
We see this with the third assertion of the CUT \texttt{testOneElement}, which holds for 6 more inputs in addition to the original one.
We aim at automatically retrieving such assertions, to harness them within a PUT.

\section{\proze}\label{sec:proze}

This section describes \proze, a technique that generates parameterized unit tests informed by runtime data.

\subsection{Overview}
\begin{figure*}
\centering
\includegraphics[scale=0.42]{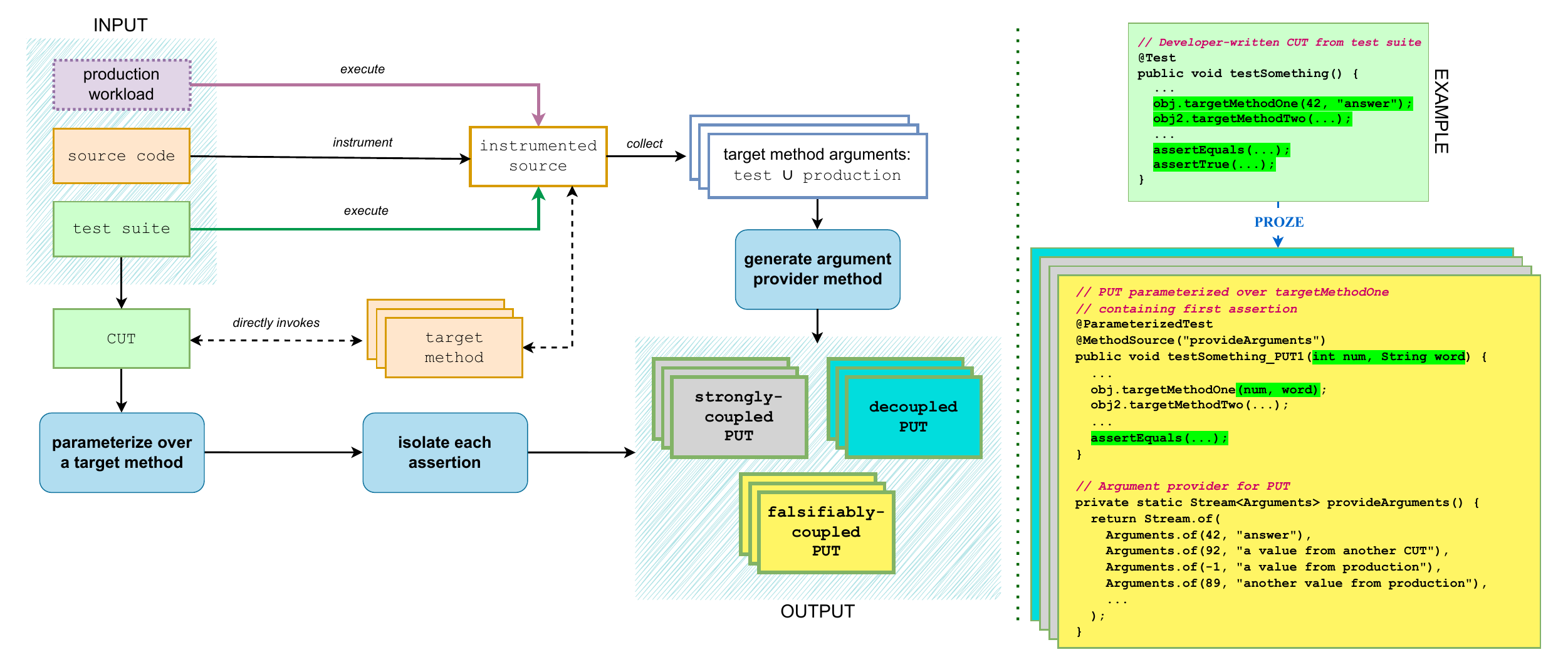}
\caption{An overview of \proze, a novel technique for the automated transformation of a developer-written conventional unit test (CUT) into parameterized unit tests (PUTs). Given a project and its original test suite composed of a set of CUTs, \proze generates a set of PUTs, which are classified as either \greyassertion, \greenassertion or \goldassertion, depending on how much the oracle is tied to the original input.
The CUT on the top right is transformed into corresponding PUTs on the bottom right.
Each PUT is fed with arguments collected at runtime with bespoke monitoring.}
\label{fig:proze-overview}
\end{figure*}

We summarize our approach in \autoref{fig:proze-overview}.
\proze accepts as input 1) the source and test code, including all developer-written CUTs, of a project, and 2) a workload for the project, which automates calling end-user functionalities (see \autoref{sec:case-studies}).
The latter is used for triggering runtime usage with realistic values. 
The final output of \proze is a set of parameterized unit tests and their corresponding argument providers.
The PUTs are classified as \greyassertion, \greenassertion, \goldassertion depending on whether the oracle in the PUT is tied to the original input, or if it holds for a larger set of inputs.
\autoref{fig:proze-overview} presents an example of this transformation for the CUT \texttt{testSomething} into a set of corresponding PUTs.
Each PUT derived from the CUT focuses on a single method and contains one of the CUT's assertions.
We highlight one of these PUTs, which contains the first assertion from the original CUT, and has parameter types \texttt{int, String}, corresponding to the parameter types of the method \texttt{targetMethodOne(int, String)}.
In the following subsections, we present details of the automated transformation of developer-written CUTs into PUTs that harness one assertion each.

\subsection{Capturing PUT inputs}\label{sec:proze-capturing-inputs}
A PUT has an argument provider, which enumerates the inputs to be used in the test.
Hence, the first step of \proze is to collect these possible inputs.
For this task, \proze sources inputs by monitoring test and production executions.
Since monitoring is costly, \proze first identifies a set of target methods \cite{kampmann2019carving} to instrument so as to monitor their invocations at runtime and capture valid arguments for them.

\emph{Selecting and instrumenting target methods}:
Given the input project, \proze statically analyzes all the CUTs within its test suite (i.e., all methods annotated with \texttt{@Test}).
As depicted in \autoref{fig:proze-overview}, \proze uses this analysis to map each CUT to the set of  methods and constructors that it directly invokes.
\proze selects \emph{target methods} that accept arguments of primitive and/or \texttt{String} types, because of implementation constraints on argument providers.
All CUTs that invoke at least one candidate target method will be tentatively transformed into PUTs by \proze.

Next, \proze instruments each target method, to capture the input arguments for each of its invocations during the execution of the program.
The instrumentation consists of instructions that trigger the serialization of the input arguments of a target method each time it gets invoked at runtime, as well as the stack trace.
These input arguments will later be used inside the argument providers of the generated PUTs.

\emph{Capturing target method arguments}: 
In order to collect a set of input arguments for each target method, the instrumented version of the program under test is executed.
\proze monitors the execution of the developer-written test suite.
This enables it to `transplant' the test inputs that appear in other tests, within the generated parameterized unit tests for the method.

Additionally, our key insight is that we can use the same instrumentation to monitor the execution of the program when it is exercised with an end-to-end workload, possibly in production. 
This allows us to collect a wide range of realistic inputs for a target method.
As illustrated in \autoref{fig:proze-overview}, \proze computes the union of all arguments passed to a target method during the execution of the test suite and during the execution of the end-to-end workload.

\subsection{Automatic generation of PUTs}\label{sec:proze-generating-puts}
At this stage, \proze generates a set of PUTs for each original CUT. 
There are three key steps that \proze takes for this transformation, presented as blue boxes in \autoref{fig:proze-overview}.

\subsubsection{Parameterize over a target method}
We design \proze to generate one test class per target method invoked by a CUT.
The values passed to the parameters of a PUT are assigned as the arguments to a single target method.
We refer to the PUT as being \emph{parameterized over the target method}.
This is illustrated in the example of \autoref{fig:proze-overview}, where the two arguments to the PUT \texttt{testSomething\_PUT1} are assigned as the arguments of \texttt{targetMethodOne}.

\subsubsection{Isolate each assertion}
A key design decision of \proze is that each PUT has one single assertion, extracted from the original CUT.
When the original CUT has multiple assertions, they verify different aspects of the expected behavior.
Isolating each  assertion within its own PUT is essential for a clear test intention through one single assertion, and  to enable individual assessment of each assertion against the captured inputs.

As a result, for a CUT that invokes $\alpha$ target methods and that specifies $\beta$ assertions, \proze generates $\alpha \times \beta$ PUTs.
Each PUT has parameters that are of the same type as the parameters of the target method it parameterizes over, and it invokes the target method with the values passed to these parameters.
We see this in the example of \autoref{fig:proze-overview}, where the PUT has parameter type \texttt{int, String}, corresponding to the target method \texttt{targetMethodOne(int, String)}, and contains the first assertion from \texttt{testSomething}.

\subsubsection{Generate argument provider method}
\proze generates one argument provider method in the test class, which is in charge of providing concrete values to the parameters of the PUTs.
For our example CUT \texttt{testSomething} of \autoref{fig:proze-overview}, \proze generates 2 test classes and 4 PUTs in total (2 target methods $times$ 2 assertions).
Specifically, for each of the target methods \texttt{targetMethodOne} and \texttt{targetMethodTwo}, \proze generates one test class with two PUTs, one for each assertion.
The PUTs parameterized over \texttt{targetMethodOne} take a pair of \texttt{int} and \texttt{String} parameters, corresponding to the parameter signature of the target method.

\proze uses the union of the arguments of a target method captured at runtime, in order to construct the argument provider.
The argument provider is implemented as a method that is linked to the PUT through a specific annotation, such as \texttt{@MethodSource} for JUnit, or with \texttt{@DataProvider} for TestNG.
As we see with the method \texttt{provideArguments} in the example of \autoref{fig:proze-overview}, this method returns an array of values, each of which corresponds to one argument that is passed to the PUT.
When the PUT is executed, this provider sequentially supplies it with one input argument.

\subsection{Classes of generated PUTs}\label{sec:proze-classification}

Once all the PUTs have been generated, \proze executes them.
As its final output, \proze categorizes each generated PUT as:
\begin{itemize}  
  \item \greyassertion if the PUT passes only when supplied with all the original arguments of the target method, and no additional arguments supplied by the provider.
  In other words, the assertion is coupled to exactly the test inputs defined in the original CUT, and it does not hold over a larger set of input arguments.
  This represents cases where the transformation of the CUT into a PUT parameterized over the target method is technically feasible, yet not useful due to the lack of an appropriate oracle.
   
  \item \greenassertion if the PUT passes for all the arguments that are supplied to it by the argument provider, including original and non-original arguments of the target method.
  This means that the assertion holds for all inputs to the target method contained in the captured union.
  The assertions that fall in this category are from two distinct categories: cases of parameterization where the oracle is indeed correct over all captured inputs (e.g., the PUT in \autoref{lst:bg-cut-put}), and cases where the assertion is not coupled to the target method's behavior.
  The latter is not interesting for testing, and the corresponding PUTs are not relevant.
  
  \item \goldassertion if the PUT passes when it is supplied with all the original arguments of the target method, as well as a subset of the additional arguments supplied by the provider.
  The latter includes target method arguments that are not present in the original CUT from which the PUT is derived.
  For a PUT that falls in this category, it is important that it fails for some of the arguments supplied by the provider, as this indicates that the assertion is sensitive to the input passed to the target method.
  Such cases represent assertions that hold over a larger set of inputs \cite{saff2008theory}, yet can successfully distinguish between valid and invalid inputs. 
  These PUTs are the most interesting, as they provide sound evidence that an existing oracle can assess the behavior of a target method over a large set of inputs.
  These PUTs can be effectively used to increase the coverage of the target's input space.
\end{itemize}

In addition to these three categories, we also find cases where the PUT fails when it is supplied with any of the original arguments of the target method, regardless of the other inputs supplied by the argument provider. 
Such ill-formed cases are due to side-effects from modifying target method invocations and removing assertion statements during PUT generation, per the core algorithm of \proze.
They are filtered out and not delivered as the output by \proze.

We note that our categorization of test assertions as strongly-coupled, falsifiably-coupled, or decoupled is novel, it captures an essential facet of the oracle problem \cite{barr2014oracle} in the context of PUT generation.

\subsection{Illustrative example}\label{sec:proze-example}
\begin{lstlisting}[style=java, showlines=true, label={lst:button-cut}, caption={\texttt{testRadioButtons} is a developer-written CUT in the \pdfboxproject project. The target method \texttt{setValue} is invoked within the test with a \texttt{String} argument (line 7). We present an excerpt of this CUT, containing two of its twelve assertions.}, float]
@Test
public void testRadioButtons() {
  File pdf1 = new File("target/RadioBtns.pdf");
  PDDocument doc1 = PDDocument.load(pdf1);
  PDAcroForm form1 = doc1 .getDocumentCatalog().getAcroForm();
  PDRadioButton radioButton1 = (PDRadioButton) form1.getField("MyRadioButton");
  radioButton1.setValue("b");
  File pdf2 = new File("target/RadioBtns-mod.pdf");
  doc1.save(pdf2);
  doc1.close();
  
  PDDocument doc2 = PDDocument.load(pdf2);
  PDAcroForm form2 = doc2 .getDocumentCatalog().getAcroForm();
  PDRadioButton radioButton2 = (PDRadioButton) form2.getField("MyRadioButton");
  assertEquals("b", radioButton2.getValue());
  assertEquals(1, radioButton2 .getSelectedExportValues().size());
  doc2.close();
}
\end{lstlisting}

\begin{lstlisting}[style=java, showlines=true, label={lst:button-put}, caption={\proze generates this parameterized unit test from the CUT \texttt{testRadioButtons} of \autoref{lst:button-cut}. It takes a \texttt{String} parameter called \texttt{value}.
The target method \texttt{setValue} is invoked within the PUT with \texttt{value} (line 8).
The PUT contains a single assertion (line 16).
The method \texttt{provideArgs} is the argument provider for the PUT, supplying 12 unique arguments to the PUT from data captured at runtime.
}, float]
@ParameterizedTest
@MethodSource("provideArgs")
public void testRadioButtons_PUT(%*\hlgreen{String value}*)) {
  File pdf1 = new File("target/RadioBtns.pdf");
  PDDocument doc1 = PDDocument.load(pdf1);
  PDAcroForm form1 = doc1 .getDocumentCatalog().getAcroForm();
  PDRadioButton radioButton1 = (PDRadioButton) form1.getField("MyRadioButton");
  radioButton1.setValue(%*\hlgreen{value}*));
  File pdf2 = new File("target/RadioBtns-mod.pdf");
  doc1.save(pdf2);
  doc1.close();
  
  PDDocument doc2 = PDDocument.load(new File("target/RadioButtons-modified.pdf"));
  PDAcroForm acroForm2 = doc2 .getDocumentCatalog().getAcroForm();
  PDRadioButton radioButton2 = (PDRadioButton) acroForm2.getField("MyRadioButton");
  assertEquals(1, radioButton2 .getSelectedExportValues().size());
  doc2.close();
}

// Provide PUT arguments from the captured union
private static Stream<Arguments> provideArgs() {
  return java.util.stream.Stream.of(
    Arguments.of("Off"),
    ...
    Arguments.of("Yes"),
    Arguments.of("b"),
    Arguments.of("c")
  );
}
\end{lstlisting}

\autoref{lst:button-cut} presents an excerpt of a CUT called \href{https://github.com/apache/pdfbox/blob/e38906b08f3eccb928c7fd7d4ab9d4213ea604bf/examples/src/test/java/org/apache/pdfbox/examples/interactive/form/TestCreateSimpleForms.java\#L219}{\texttt{testRadioButtons}}, written by the developers of \pdfboxproject, a notable open-source PDF manipulation library \cite{butler2020maintaining}.
The test loads a PDF document which contains a form, and fetches the form field containing a group of radio buttons (lines 3 to 6).
The radio button field can be set to one of three options, \texttt{"a"}, \texttt{"b"}, or \texttt{"c"}.
Next, on line 7, the method \texttt{setValue} is invoked with the \texttt{String} value \texttt{"b"}, to select the option \texttt{"b"} of the button, also updating its visual appearance.
This modified PDF is saved and closed (lines 8 to 10), and then reloaded as a new document (line 12).
After fetching the same radio button field, the two assertions verify that the selected value of the button is still \texttt{"b"} (line 15), and that there is only one \href{https://www.pdfscripting.com/public/Checkboxes-and-Radio-Buttons.cfm}{export value} corresponding to the selected value (line 16).

\proze statically analyzes \texttt{testRadioButtons}, identifies the method \texttt{setValue} as a target method, and consequently instruments this method to capture its \texttt{String} arguments at runtime.
During the execution of the test suite of \pdfboxproject, as well as its execution under a production workload, \proze monitors the invocations of \texttt{setValue}, capturing the 12 unique \texttt{String} arguments that \texttt{setValue} gets invoked with.
This set of arguments contains the \texttt{String} \texttt{"b"} that \texttt{setValue} is invoked with in the CUT.
We refer to \texttt{"b"} as the \emph{original argument} of \texttt{setValue} within the CUT.

After capturing this union of input arguments at runtime, \proze generates two PUTs for \texttt{testRadioButtons} that are parameterized over \texttt{setValue}, one for each of the two assertions in \autoref{lst:button-cut}.
Lines 1 to 18 of \autoref{lst:button-put} present the PUT that includes the second assertion of the CUT (line 16 of \autoref{lst:button-cut}).
The PUT is annotated with the JUnit5 \texttt{@ParameterizedTest} annotation.
This PUT has a single \texttt{String} parameter corresponding to the \texttt{String} parameter type of \texttt{setValue},  highlighted on line 3, and it contains one assertion at line 16.
\proze encapsulates the 12 captured \texttt{String} arguments of \texttt{setValue} into the data provider \texttt{provideArgs} (lines 20 to 29 in \autoref{lst:button-put}).
This data provider is linked to the generated PUT through the \texttt{@MethodSource} JUnit5 annotation (line 2).
When executed, the PUT accepts each argument sequentially from the provider and assigns it as the argument for the invocation of \texttt{setValue} (highlighted on line 8).
This results in 12 unique executions of the test, each with a distinct argument supplied to the PUT by \texttt{provideArgs}.
\proze classifies this PUT as \goldassertion with \texttt{setValue}, since it passes for 2 of the 12 arguments it is invoked with, i.e., the original argument \texttt{"b"} as well as one additional argument \texttt{"c"}.
On the other hand, the generated PUT containing the assertion on line 15 of \autoref{lst:button-cut} is classified as \greyassertion, as it passes only with \texttt{"b"}.

\subsection{Implementation details}\label{sec:proze-implementation}
\proze is implemented in Java.
It is implemented using open-source libraries, including Spoon \cite{pawlak:hal-01169705} for code analysis and generation, and 
Glowroot \cite{tiwari2022Pankti} for dynamic instrumentation and monitoring.
\proze generates PUTs in two of the most widely used test frameworks in Java \cite{kim2021studying}, JUnit5 and TestNG. 
Our implementation is publicly available at \url{https://github.com/ASSERT-KTH/proze/}.

\section{Experimental Methodology}
This section describes our methodology to assess the ability of \proze to transform developer-written CUTs into PUTs, and in particular to identify the assertions that are suitable for parameterization.

\subsection{Research questions}\label{sec:rqs}
We answer the following research questions as part of our evaluation.
\begin{enumerate}
    \item [RQ1] [\emph{PUT arguments}]: To what extent does \proze collect and increase the input space coverage for the target methods?
    \item [RQ2] [\emph{PUT classification}]: To what extent does \proze extend the range of the oracle over the input space?
    \item [RQ3] [\emph{PUT representativeness}]: 
    How do the PUTs generated by \proze relate to the testing practices of developers?
\end{enumerate}

\subsection{Dataset: Applications and workloads}\label{sec:case-studies}
\autoref{tab:case-studies} presents details of the five Java modules we use for the evaluation of our approach, including the link to the exact version we use for our experiments (Commit \textsc{SHA}), and the number of lines of Java source code (\textsc{LoC}, counted with \texttt{cloc}).
We also report the number of developer-written CUTs in their test suite (\textsc{Total CUTs}), and the \textsc{Total Assertions} across these CUTs.

\begin{table}[htbp]
\centering
\caption{Description of the five Java modules used as case studies in our evaluation.}\label{tab:case-studies}

\begin{tabular}{lrrrr}
\toprule
\textsc{Module} &
\textsc{SHA} &
\textsc{LoC} &
{\begin{tabular}[c]{@{}r@{}}\textsc{Total} \\ \textsc{CUTs} \end{tabular}} &
{\begin{tabular}[c]{@{}r@{}}\textsc{Total} \\ \textsc{Assertions} \end{tabular}}
\\
\midrule
\fontbox & \href{https://github.com/apache/pdfbox/tree/2.0.24/fontbox}{\texttt{8876e8e}} & 16,815 & 25 & 156 \\
\xmpbox  & \href{https://github.com/apache/pdfbox/tree/2.0.24/xmpbox}{\texttt{8876e8e}} & 7,181 & 54 & 222 \\
\pdfbox & \href{https://github.com/apache/pdfbox/tree/2.0.24/pdfbox}{\texttt{8876e8e}} & 84,764 & 429 & 1,633 \\
\query & \href{https://github.com/wso2-extensions/identity-inbound-auth-saml/tree/v5.11.41/components/org.wso2.carbon.identity.query.saml}{\texttt{3ea38dd}} & 2,472 & 33 & 42 \\
\sso & \href{https://github.com/wso2-extensions/identity-inbound-auth-saml/tree/v5.11.41/components/org.wso2.carbon.identity.sso.saml}{\texttt{3ea38dd}} & 13,520 & 83 & 152 \\
\midrule
\textbf{\textsc{Total}} & &  & \textbf{624} & \textbf{2,205} \\
\bottomrule
\end{tabular}
\end{table}
 
\textbf{\pdfboxproject}:
The first three modules we consider belong to \href{https://pdfbox.apache.org/}{\pdfboxproject}, a popular open-source project by the Apache Software Foundation, which supports operations for working with PDF documents \cite{butler2020maintaining}.
We experiment with three modules of \pdfboxproject version 2.0.24, called \fontbox, \xmpbox, and \pdfbox.
The \fontbox module (first row of \autoref{tab:case-studies}) handles fonts within PDF documents.
It has more than 16K lines of Java source code, and its test suite contains 25 developer-written CUTs implemented as JUnit tests.
The total number of assertions across these 25 CUTs is 156.
The second module, \xmpbox, is responsible for managing metadata within PDF documents, based on the eXtensible Metadata Platform \href{https://www.osplabs.com/wp-content/uploads/whitepaper/xmp_whitepaper.pdf}{(XMP)} specification.
Its source code comprises of 7K lines of Java code, and the 54 CUTs in its test suite contain 222 assertions in total.
The core module of the \pdfboxproject project is \pdfbox, which is the third one we experiment with.
With upwards of 84K lines of Java source code, \pdfbox is the largest module in our dataset.
Its test suite has 429 developer-written CUTs containing 1,633 assertion statements in total.

We run the test suite of \pdfbox in order to capture the arguments with which its target methods are invoked at runtime.
We also exercise \pdfbox with the production workload from \cite{tiwari2022Pankti}, performing operations that are typical of PDF documents, including encryption, decryption, and conversion to/from text and images, on 5 real-world PDF documents sourced from \cite{garfinkel2009bringing}.

\textbf{\saml}:
The second project we consider for our evaluation is \texttt{identity-inbound-auth-saml} (henceforth referred to as \saml).
It is an extension of the \href{https://wso2.com/identity-and-access-management/}{WSO2 identity server}, which provides identity and access management across commercial applications worldwide.
The \saml project contains functionalities for performing user authentication using the Security Assertion Markup Language (SAML) protocol \cite{schwartz2018saml}.
We work with two modules within \saml version 5.11.41.
The first module, \query (row 4 of \autoref{tab:case-studies}), handles application authentication- and authorization-related queries.
It has nearly 2.5K lines of Java code, and 33 developer-written TestNG CUTs.
The \sso module supports Single SignOn (SSO) for SAML.
There are 13.5K lines of Java code within this module, and 83 CUTs with 152 assertions.

\proze instruments each target method called by the CUTs across \query and \sso with the goal of capturing their input arguments.
These arguments are captured while the test suite of \saml is run and when the field workload is executed.
For this experiment, we use three \href{https://github.com/wso2/samples-is/releases}{sample} web applications for \saml: \texttt{pickup-dispatch}, \texttt{pickup-manager}, and \texttt{travelocity}.
We first deploy \saml and register the three sample applications using the GUI of the WSO2 identity server.
Then we configure SSO and Single LogOut (SLO) for all three applications.
We exercise SSO by logging into \texttt{pickup-dispatch} and consequently accessing \texttt{pickup-manager} and \texttt{travelocity}.
Similarly, we perform SLO by logging out from one application.
We also modify the configuration of the applications, such as sharing user attributes (i.e., \emph{claims} in SAML), including name, country, and email, and enabling claim-sharing between the three applications.

\subsection{Protocol}\label{sec:protocol}
We run each of the five modules per their test and field workloads detailed in \autoref{sec:case-studies}, while \proze captures arguments for the target methods it identifies within their developer-written CUTs.
For a target method, if \proze captures more than one argument from different execution contexts, i.e., test and field, it generates PUTs by transforming each CUT that invokes the method, as well as the corresponding argument providers.
Per \autoref{sec:proze}, each generated PUT is parameterized over a single target method, and contains one of the CUT's assertions.
Based on this methodology, this subsection presents the protocol we employ to answer the three research questions listed in \autoref{sec:rqs}.

\emph{RQ1 [PUT arguments]}:
As the answer to our first question, we report the number of target methods for which \proze successfully captures a union of arguments at runtime over test and field executions, and transforms the CUTs that invoke them into PUTs.
As argued by Kuhn \etal \cite{kuhn2020inputcoverage}, adequately covering the input space within tests has a positive relationship with test quality.
Therefore, for each target method, the key metric for this RQ is the number of unique arguments \proze captures for it.
These arguments are encapsulated within the argument provider for each PUT that is parameterized over the target method.
This set of arguments represents the increase in the input space coverage of the target method. 

\emph{RQ2 [PUT classification]}:
\proze outputs a set of PUTs which are classified as either \greyassertion, \greenassertion or \goldassertion. 
To answer RQ2, we report the number of PUTs in each of these categories. 
Per \autoref{sec:proze-classification}, a falsifiably-coupled PUT provides concrete evidence that its oracle is valid over a larger range of test inputs than the original input of the CUT from which it is derived.

\emph{RQ3 [PUT representativeness]}:
For this RQ, we present a qualitative analysis of the three most notable techniques for the automatic generation of PUTs, with respect to the PUTs generated by \proze.
The three related works we consider for this analysis, \cite{fraser2011generating},  \cite{kampmann2019carving}, and \cite{xu2024mr} employ diverse methodologies for automatically generating PUTs.
We summarize their approaches and reason about the representativeness of the PUTs generated by them against those generated by \proze.

\section{Experimental Results}
This section presents the answers to the research questions introduced in \autoref{sec:rqs}, based on our experiments with the \nbmodules modules described in \autoref{sec:case-studies}, and using the protocol detailed in \autoref{sec:protocol}.

\subsection{RQ1: PUT arguments}

\begin{table}[t]
\centering
\caption{RQ1: \proze captures runtime arguments (\textsc{\#Cap. Args}) for \textsc{Target methods} across the \nbmodules \textsc{Module}s, generating \textsc{PUTs} from the \textsc{CUTs} that directly invoke them with \texttt{\#Orig. Args}.}\label{tab:rq1}
\resizebox{\columnwidth}{!}{
\begin{tabular}{lrrrrrr}
\toprule
\multicolumn{2}{l}{\textsc{Module}} & 
\textsc{Target Methods} & 
\textsc{CUTs} & 
{\begin{tabular}[c]{@{}r@{}}\textsc{\#Orig.} \\ \textsc{Args} \end{tabular}} &
\textsc{PUTs} & 
{\begin{tabular}[c]{@{}r@{}}\textsc{\#Cap.} \\ \textsc{Args} \end{tabular}}
\\
\toprule
\rowcolor[rgb]{0.827, 0.906, 0.941} 
\multicolumn{2}{l}{\textbf{\fontbox}} & 
\textbf{12} & 
\textbf{15} & 
\textbf{med. 2} &
\textbf{188} & 
\textbf{med. 60} \\ 
\multicolumn{1}{l|}{\multirow{3}{*}{\textsc{examples}}} & \multicolumn{2}{r}{\texttt{getLeftSideBearing}} & 2 & 8 & 15 & 205 \\
\multicolumn{1}{l|}{} & \multicolumn{2}{r}{\texttt{getGlyph}} &  3 & 5 & 14 & 187 \\ 
\multicolumn{1}{l|}{} & \multicolumn{2}{r}{\texttt{getGlyphId}} & 1 & 2 & 8 & 68 \\
\midrule
\rowcolor[rgb]{0.827, 0.906, 0.941} 
\multicolumn{2}{l}{\textbf{\xmpbox}} & 
\textbf{27} & 
\textbf{36} & 
\textbf{med. 1} &
\textbf{594} & 
\textbf{med. 5} \\ 
\multicolumn{1}{l|}{\multirow{3}{*}{\textsc{examples}}} & \multicolumn{2}{r}{\texttt{createText}} & 6 & 7 & 32 & 738 \\
\multicolumn{1}{l|}{} & \multicolumn{2}{r}{\texttt{addQualifiedBagValue}} & 2 & 3 & 10 & 630 \\ 
\multicolumn{1}{l|}{} &
\multicolumn{2}{r}{\texttt{addUnqualifiedSeqValue}} &
3 & 4 & 30 & 117 \\
\midrule
\rowcolor[rgb]{0.827, 0.906, 0.941} 
\multicolumn{2}{l}{\textbf{\pdfbox}} & 
\textbf{55} & 
\textbf{155} & 
\textbf{med. 1} & 
\textbf{1,388} & 
\textbf{med. 8} \\ 
\multicolumn{1}{l|}{\multirow{3}{*}{\textsc{examples}}} & \multicolumn{2}{r}{\texttt{encode}} & 2 & 2 & 12 & 2,694 \\
\multicolumn{1}{l|}{} & \multicolumn{2}{r}{\texttt{getStringWidth}} & 2 & 3 & 6 & 2,351 \\ 
\multicolumn{1}{l|}{} & \multicolumn{2}{r}{\texttt{getPDFName}} & 13 & 1,016 & 64 & 1,193 \\
\midrule
\rowcolor[rgb]{0.827, 0.906, 0.941} 
\multicolumn{2}{l}{\textbf{\query}} & 
\textbf{\nbquerytargetMethods} & 
\textbf{\nbquerycuts} &
\textbf{med. 1} & 
\textbf{\nbqueryputs} & 
\textbf{med. 3} \\
\multicolumn{1}{l|}{\multirow{2}{*}{\textsc{examples}}} & \multicolumn{2}{r}{\texttt{getServiceProviderConfig}} & 1 & 1 & 3 & 3 \\
\multicolumn{1}{l|}{} & \multicolumn{2}{r}{\texttt{getIssuer}} & 2 & 2 & 5 & 3 \\ 
\midrule
\rowcolor[rgb]{0.827, 0.906, 0.941} 
\multicolumn{2}{l}{\textbf{\sso}} & 
\textbf{\nbssotargetMethods} & 
\textbf{\nbssocuts} & 
\textbf{med. 1} & 
\textbf{\nbssoputs} & 
\textbf{med. 6} \\ 
\multicolumn{1}{l|}{\multirow{3}{*}{\textsc{examples}}} & \multicolumn{2}{r}{\texttt{unmarshall}} & 2 & 2 & 3 & 35 \\
\multicolumn{1}{l|}{} & \multicolumn{2}{r}{\texttt{decodeForPost}} & 1 & 1 & 1 & 24 \\
\multicolumn{1}{l|}{} & \multicolumn{2}{r}{\texttt{encode}} & 1 & 1 & 2 & 23 \\ 
\midrule
\rowcolor[rgb]{0.827, 0.906, 0.941} 
\multicolumn{2}{l}{\textbf{\textsc{Total}}} & 
\textbf{\nbtargets} & 
\textbf{\nbcuts} & 
\textbf{2,739} & 
\textbf{\nbputs} & 
\textbf{13,021} \\
\bottomrule
\end{tabular}}
\end{table}

\autoref{tab:rq1} summarizes the results for our first RQ.
For each \textsc{Module}, the highlighted rows present the total number of \textsc{Target methods}, the total number of \textsc{CUTs} that directly invoke them, and the median number of original arguments passed to these methods from the \textsc{CUTs} (\textsc{\#Orig. Args}).
Next, we give the number of \textsc{PUTs} that \proze generates from these CUTs, and the median number of arguments captured (\textsc{\#Cap. Args}) at runtime for the target methods in the module.
Under each module, we list examples of target methods for which \proze captures a large number of arguments, including the number of CUTs that invoke them, the number of original arguments that they use to do so, the number of PUTs generated by \proze for the target method, as well as the number of arguments captured for it.

Per the methodology described in \autoref{sec:proze}, \proze statically analyzes each of the 624 CUTs across the \nbmodules modules in \autoref{tab:case-studies}, to select target methods, and instruments these methods to capture their arguments at runtime.
However, not all of these 624 CUTs can be transformed into PUTs.
First, not all of them contain appropriate target methods per the criteria explained in \autoref{sec:proze-capturing-inputs}.
Second, even if a CUT contains a target method, the method may not be invoked at runtime.
Overall, we see that \proze captures data for \nbtargets target methods across test and field executions, which are directly invoked by 248 of the 624 CUTs in the \nbmodules modules.
This means that these \nbtargets target methods are invoked in the field and/or by at least one developer-written test.
Furthermore, for each of these \nbtargets methods, \proze captures at least one runtime argument in addition to the original argument, from different sources.
In total, \proze captures 13,021 arguments at runtime for the \nbtargets target methods, which are passed to \nbputs generated PUTs.

The first few rows of \autoref{tab:rq1} show the results for the three \pdfboxproject modules, \fontbox, \xmpbox, and \pdfbox.
\proze generates \nbpdfputs PUTs in total for the \nbpdfTargetMethods target methods across these three modules.
The PUTs are derived from the \nbpdfcuts CUTs that invoke the \nbpdfTargetMethods methods directly.
We see the largest numbers for the \pdfbox module, owing to its large size, as highlighted in \autoref{tab:case-studies}.
The 55 target methods in \pdfbox are invoked by 155 CUTs, from which \proze generates 1,388 PUTs.
The median number of original arguments passed to these methods from the CUTs is 1, whereas the median number of arguments captured by \proze for these 55 methods is 8.
This means that \proze increases the coverage of the input space of 22 methods only by a few values.
For example, one PUT invokes the method \texttt{setFlyScale(float)} with the original argument passed by the CUT \texttt{saveAndReadTransitions}, and one more argument captured when running the whole test suite and workload.
On the other hand, \proze significantly increases the input space coverage for the other 23 methods.
For example, for the method \texttt{encode(int)}, \proze captures 2,694 arguments, the maximum among the target methods in our dataset.
There are two CUTs that directly invoke \texttt{encode} with one original argument each. 
\proze generates 12 PUTs from these 2 CUTs.
Each PUT takes an \texttt{int} parameter that is supplied values from the argument provider method, which includes these 2,694 captured arguments.
\revised{\proze augments the input space coverage of \texttt{encode} by three orders of magnitude.}

The results obtained for the two modules, \query and \sso, of \saml are given in the next set of rows in \autoref{tab:rq1}. 
Within \saml, the maximum number of arguments is observed for the target method \texttt{unmarshall} which is called directly by 2 CUTs \texttt{testMarshall} and \texttt{testUnmarshall}, with 1 original argument each.
\proze generates 3 PUTs for \texttt{umarshall}, along with a data provider that generates 35 stringified XMLs. 
Overall, \proze generates 8 PUTs for the \query module and 109 PUTs for the \sso module, with their argument providers having a median of 3 and 6 arguments, respectively.

\begin{mdframed}[nobreak=true,style=mpdframe, frametitle=RQ1: PUT arguments]
\proze captures 13,021 arguments at runtime for the \nbtargets methods across our five study subjects.
These  arguments are encapsulated within argument provider methods for the \nbputs generated PUTs.
Invoking the target methods with these arguments within the PUTs increases their input space coverage, sometimes up to several orders of magnitude.
\end{mdframed}

\subsection{RQ2: PUT classification}
To answer this RQ, we execute each of the \nbputs PUTs generated by \proze, and delve into our PUT classification protocol described in \autoref{sec:protocol}.
For the five modules, \autoref{tab:rq2} presents the number of PUTs that pass at least with the original arguments, and fall into each of the three categories, \greyassertion, \goldassertion, and \greenassertion.
We disregard \nbreds ill-formed PUTs per \autoref{sec:proze-classification} for this RQ. 

\begin{table}[htbp]
\centering
\caption{RQ2: PUT classification as \textsc{Strongly-coupled}, \textsc{Falsifiably-coupled}, or \textsc{Decoupled} with the target method invocation.}\label{tab:rq2}
\begin{tabular}{lr
> {\columncolor[HTML]{d3d3d3}}r
> {\columncolor[HTML]{fff564}}r
> {\columncolor[HTML]{01dddd}}r
}
\toprule
\textsc{Module} &
\textsc{PUTs} &
{\begin{tabular}[c]{@{}r@{}}\textsc{Strongly} \\ \textsc{-coupled} \end{tabular}} &
{\begin{tabular}[c]{@{}r@{}}\textsc{Falsifiably} \\ \textsc{-coupled} \end{tabular}} &
\textsc{Decoupled} \\
\midrule
\texttt{\fontbox} & 135 & 31 & 12 & 92 \\
\texttt{\xmpbox}  & 543  & 94 & 40 & 409 \\
\texttt{\pdfbox} & 1,150 & 510 & 161 & 479 \\
\texttt{\query} & 7 & \nbquerygreys & \nbquerygolds & \nbquerygreens \\
\texttt{\sso} & 79 & \nbssogreys & \nbssogolds & \nbssogreens \\
\midrule
\textbf{\textsc{Total}} & \textbf{1,914} & \textbf{\nbgreys} & \textbf{\nbgolds} & \textbf{\nbgreens} \\
\bottomrule
\end{tabular}
\end{table}

\begin{lstlisting}[style=java, showlines=true, label={lst:rq2-cut-put}, caption={The CUT \texttt{testBagManagement} in \xmpbox has 4 assertions, and calls target method \texttt{createText}. \proze captures 738 arguments for \texttt{createText}, and generates the argument provider method \texttt{provideCreateTextArgs}. 
This provider supplies arguments to 4 generated PUTs parameterized over \texttt{createText}. The parts  common to the 4 generated PUTs are included in \texttt{testBagManagement\_PUT\_N}.}, float]
@Test
public void testBagManagement() {
  XMPMetadata parent = XMPMetadata.createXMPMetadata();
  XMPSchema schem = new XMPSchema(parent, "nsURI", "nsSchem");
  String bagName = "BAGTEST";
  String value1 = "valueOne";
  String value2 = "valueTwo";
  schem.addBagValue(bagName, schem.getMetadata().getTypeMapping() .createText(null, "rdf", "li", value1));
  schem.addQualifiedBagValue(bagName, value2);
  List<String> values = schem.getUnqualifiedBagValueList(bagName);
  assertEquals(value1, values.get(0));
  assertEquals(value2, values.get(1));
  schem.removeUnqualifiedBagValue(bagName, value1);
  List<String> values2 = schem.getUnqualifiedBagValueList(bagName);
  assertEquals(1, values2.size());
  assertEquals(value2, values2.get(0));
}%*\par\noindent\dotfill*)

// Generated PUT parameterized over createText 
@ParameterizedTest
@MethodSource("provideCreateTextArgs")
public void testBagManagement_PUT_N(String ns, String prefix, String propName, String value) {
  ...
  schem.addBagValue(bagName, schem.getMetadata().getTypeMapping() .createText(ns, prefix, propName, value));
  ...
  // one of lines 11, 12, 15, or 16 per PUT
  ...
}

// Argument provider with 738 captured inputs
static Stream<Arguments> provideCreateTextArgs() {
  return java.util.stream.Stream.of(
    Arguments.of(null, "rdf", "li", "valueOne"),
    ...
    Arguments.of(null, "pdf", "PDFVersion", "1.4"),
    Arguments.of("nsURI", "nsSchem", "li", "valueTwo")
  );
}
\end{lstlisting}

Let us consider a CUT from \xmpbox called \href{https://github.com/apache/pdfbox/blob/8876e8e1a0adbf619cef4638cc3cea073e3ca484/xmpbox/src/test/java/org/apache/xmpbox/schema/XMPSchemaTest.java\#L62}{\texttt{testBagManagement}} (lines 1 to 17 of \autoref{lst:rq2-cut-put}).
This CUT invokes the target method \texttt{createText} with an argument containing 4 \texttt{String} values (line 8), and contains 4 assertions (lines 11, 12, 15, and 16).
The original argument for \texttt{createText} within this CUT is \texttt{null, "rdf", "li", "valueOne"}.
In addition to this original argument, \proze captures 737 unique arguments with which \texttt{createText} is invoked at runtime.
Using this union of captured arguments, \proze generates the argument provider method \texttt{provideCreateTextArgs} (lines 31 to 39 in \autoref{lst:rq2-cut-put}).
Furthermore, \proze generates 4 PUTs for \texttt{testBagManagement}, one for each of its 4 assertions.
Since the PUTs are parameterized over \texttt{createText}, they accept 4 \texttt{String} values (line 23) from the argument provider method.
For brevity, we showcase an excerpt that is common to these 4 PUTs (lines 20 to 29), specifically the invocation of \texttt{createText} with the PUT parameters on line 25.
Note that each of the 4 generated PUTs contains a single assertion from the original CUT.

\noindent\emph{\greyassertion cases}: The two PUTs containing the first and third assertion of \texttt{testBagManagement} (lines 11 and 15, respectively in \autoref{lst:rq2-cut-put}) only pass when the argument they receive from \texttt{provideCreateTextArgs} is the original argument (on line 34).
We classify them as strongly-coupled to the original argument of \texttt{createText}.
Per \autoref{tab:rq2}, we find \nbgreys such PUTs for which the original argument is critical for the assertion to pass.
In other words, the oracle is completely bound to the original test input.

\noindent\emph{\greenassertion case}: Next, the PUT containing the assertion on line 12 of \texttt{testBagManagement} holds over all 738 arguments provided by \texttt{provideCreateTextArgs} in the generated parameterized unit test.
We categorize this PUT as decoupled from the argument of \texttt{createText}, as the invocation of \texttt{createText} with all arguments pass.
This means \proze is not able to prove that the argument propagates to the assertion, and that the oracle is coupled to the PUT inputs.
In \autoref{tab:rq2}, we report \nbgreens such PUTs that contain an assertion that is likely independent of the arguments passed to the target method.
To be conservative, we assume that all of them are in this situation, but we note that it is theoretically possible that the assertion still verifies an interesting behavior over all inputs passed to the PUT (cf. \autoref{lst:bg-cut-put}).

\noindent\emph{\goldassertion case}: Finally, the PUT containing the assertion on line 16 of \texttt{testBagManagement} is valid for the original argument, and for one additional argument, \texttt{"nsURI", "nsSchem", "li", "valueTwo"} (line 37 of \autoref{lst:rq2-cut-put}), but invalid for the other ones.
This makes this PUT falsifiably-coupled to the target method \texttt{createText}, with a proof that the oracle is coupled to the PUT input.
There are \nbgolds such PUTs for which \proze has discovered at least one additional valid input by monitoring runtime executions.
We find the maximum increase in input space coverage for the target method \texttt{encode} in \pdfbox, for which \proze generates 12 PUTs (\autoref{tab:rq1}), of which 10 are falsifiably-coupled to the arguments of \texttt{encode}, each passing for 2,692 of the 2,694 captured inputs. 

Falsifiably-coupled PUTs are ready to be turned into PUTs that can be added in the original test suite.
These PUTs include an assertion that demonstrably assesses the behavior of the target method, with an argument provider that feeds realistic values to the PUT.
In \proze, the final steps to make a falsifiably-coupled PUT ready to be incorporated into the suite are: (i) limit the provider to deliver only valid arguments, by removing the falsifying arguments from the initial provider; (ii) if multiple PUTs use the same argument providers, merge the assertions of these different PUTs into one PUT.


\begin{mdframed}[nobreak=true,style=mpdframe, frametitle=RQ2: PUT classification]
\proze generates \nbgolds PUTs that include an assertion that is falsifiably-coupled to the arguments supplied to a target method, i.e., it is proven to successfully discriminate the behavior of the target method depending on the arguments passed to the PUT.
To our knowledge, \proze is the first technique that is able to generate PUTs using realistic inputs with the guarantee of a valid oracle.
\end{mdframed}

\subsection{RQ3: Test representativeness}\label{sec:rq3}

\begin{table*}[]
\centering
\caption{RQ3: Qualitative analysis of \proze in the testing process, compared to three previous studies on automated PUT generation}\label{tab:rq3}
\begin{tabular}{r|r|r|r|
> {\columncolor[HTML]{CFF0FE}}r
}
\toprule
& \textsc{Fraser \& Zeller}, 2011 \cite{fraser2011generating} &
\textsc{Basilisk}, 2019 \cite{kampmann2019carving} & 
\textsc{MR-Scout}, 2024 \cite{xu2024mr} & 
\proze \\
\toprule
\textsc{Language} & Java & C & Java & Java \\
\midrule
\textsc{Input} &
source class & 
\begin{tabular}[c]{@{}r@{}}manually-written system tests \\ + system tests generated \\ with fuzzing\end{tabular} & 
\begin{tabular}[c]{@{}r@{}}developer-written \\ test suite \end{tabular} & 
\begin{tabular}[c]{@{}r@{}}developer-written test \\ suite + workload \end{tabular} \\
\midrule
\begin{tabular}[c]{@{}r@{}}\textsc{Test}\\ \textsc{Inputs}\end{tabular} &
\begin{tabular}[c]{@{}r@{}}symbolic preconditions\\ through mutation\end{tabular} &
\begin{tabular}[c]{@{}r@{}}random values from\\ fuzzing system test inputs\end{tabular} &
\begin{tabular}[c]{@{}r@{}}instances of class under test \\ through search-based approach \end{tabular} &
\begin{tabular}[c]{@{}r@{}}union of arguments from\\ test and field invocations\end{tabular} \\
\midrule
\begin{tabular}[c]{@{}r@{}}\textsc{Test}\\ \textsc{Oracle}\end{tabular} &
symbolic postconditions & 
\begin{tabular}[c]{@{}r@{}}implicit assertion (failure),\\ increase in branch coverage\end{tabular} & 
\begin{tabular}[c]{@{}r@{}}assertion with\\ metamorphic relation (MR)\end{tabular} 
& developer-written assertion \\
\midrule
\begin{tabular}[c]{@{}r@{}}\textsc{Test}\\ \textsc{Scenario}\end{tabular} &
\begin{tabular}[c]{@{}r@{}}optimized subset of relevant\\ assumptions, sequence of\\  method calls, assertions\end{tabular} &  
carved from system test & 
\begin{tabular}[c]{@{}r@{}}multiple invocations of \\ methods from the same class \\ with different arguments \end{tabular} &
\begin{tabular}[c]{@{}r@{}}derived from \\ developer-written CUT\end{tabular} \\
\midrule
\begin{tabular}[c]{@{}r@{}}\textsc{Test}\\ \textsc{Framework}\end{tabular} &
\texttt{assume}, \texttt{assert} clauses & 
functions in C & 
\begin{tabular}[c]{@{}r@{}}parameterized methods\\ called by JUnit5 tests \end{tabular} &
\begin{tabular}[c]{@{}r@{}}full-fledged JUnit5 \\ + TestNG tests \end{tabular} \\
\midrule
\textsc{Example} &
Section 5.4 of \cite{fraser2011generating} &
running example in \cite{kampmann2019carving} & 
Figure 4 of \cite{xu2024mr} & 
\autoref{lst:button-put}, \autoref{lst:rq2-cut-put} \\
\bottomrule
\end{tabular}
\end{table*}

\autoref{tab:rq3} summarizes our insights from the analysis of the PUTs generated by \proze with respect to those generated by \cite{fraser2011generating}, \cite{kampmann2019carving}, and \cite{xu2024mr}.

With the exception of \textsc{Basilisk} \cite{kampmann2019carving} which focuses on C programs, all the studies do PUT generation for Java projects.
Per the \textsc{Input} row, \proze and \textsc{MR-Scout} \cite{xu2024mr} are the only techniques that use developer-written CUTs from the existing test suite of a project as the seed for their PUT generation efforts.
This guarantees that developer intentions propagate to the generated PUTs, making them understandable for developers.
The \textsc{Test Inputs} row presents the methodology employed by each technique to provide multiple inputs to the PUTs.
The PUTs generated by Fraser and Zeller \cite{fraser2011generating} include symbolically generated inputs, while \textsc{Basilisk} uses random inputs generated from fuzzing, and \textsc{MR-Scout} uses inputs generated through search-based techniques.
\proze is novel in this respect, as it captures actual usages of target methods during test and field executions.
PUTs have previously not been automatically generated with realistic test inputs collected at runtime.
The following row presents the kind of \textsc{Test Oracle} included in the generated tests.
\textsc{MR-Scout} and \proze are the only techniques that reuse developer-written assertions in the generated PUTs.
Preserving the developers' assertions as part of the generated PUTs contributes to their readability with respect to the test intention.
Likewise, the tests generated by \textsc{MR-Scout} and \proze inherit \textsc{Test Scenario}s from original developer-written CUTs.
The goal of a codified MR generated by \textsc{MR-Scout} is to express a metamorphic relation that exists between the initial and subsequent states of an object.
On the other hand, within a \proze PUT, an existing assertion is evaluated against a larger set of inputs than in the original CUT.
The concept of extending the range of an oracle over more inputs within a PUT is both new and powerful. 
\proze is also the only technique to generate PUTs that are native JUnit5 (\autoref{lst:button-put} and \autoref{lst:rq2-cut-put}) and TestNG parameterized tests, the two most popular \textsc{Test Framework}s in Java \cite{kim2021studying}.
The invocation of the PUT with distinct arguments is fully compatible with existing test runners.
Compared to the three most related techniques for PUT generation, we argue that the PUTs generated by \proze are more representative of real-world tests and testing practices.
First, \proze transforms a developer-written CUT into a PUT for a target method, which guarantees that the PUT represents a real-world test scenario.
Second, the inputs to the PUT are concrete values captured at runtime, and supplied to the PUT explicitly through an argument provider method.
Finally, the assertion in the PUT is not implicit or inferred, but explicit and adopted directly from the developer-written CUT, capturing a clear and scoped test intention.

\begin{mdframed}[nobreak=true,style=mpdframe, frametitle=RQ3: PUT representativeness]
\proze derives PUTs from developer-written CUTs, which guarantees that the PUTs have a clear test intention.
\proze exploits best-in-class testing frameworks, and the generated tests are representative of current testing practices.
The \proze PUTs are the first ever to be informed by runtime data from test and field executions.
\end{mdframed}

\subsection{Threats to validity}

\revised{
A threat to the internal validity of our evaluation arises from the implementation of \proze, which may contain bugs.
The limited number of study subjects [13] in our evaluation has implications on the external validity of our findings.
}

\section{Related Work}
Since they were first proposed by Tillmann and Schulte \cite{tillmann2005parameterized}, PUTs have received considerable attention in the literature.
The empirical study on open-source .NET projects by Lam \etal \cite{lam2018characteristic} identifies common implementation patterns and code smells in PUTs written with the PEX framework \cite{tillmann2008pex}.
Several studies have also proposed techniques for automatically generating PUTs.
UnitMeister and AxiomMeister \cite{tillmann2006discovering} use symbolic execution techniques for generalizing CUTs in order to produce a new set of PUTs \cite{tillmann2006unit}.
The technique proposed by Thummalapenta and colleagues \cite{thummalapenta2011retrofitting} for retrofitting CUTs into PUTs requires significant manual effort to create generalized oracles that are valid over all inputs.
Compared to these studies, \proze is fully automated, it does not require manual intervention to generalize the oracle or to generate PUTs.  

Tsukamoto \etal \cite{tsukamoto2018autoput} propose AutoPUT to automate the retrofitting of PUTs from developer-written CUTs.
AutoPUT identifies a set of CUTs with similar Abstract Syntax Trees (ASTs) and consolidates them into a single PUT implemented as JUnit4 \texttt{@Theory} methods.
A similar approach is employed by Azamnouri \etal \cite{azamnouri2021compressing} to convert CUTs into PUTs as a means of compressing the test suite.
Menéndez \etal \cite{menendez2021diversifying} propose the concept of diversified focused testing, which involves generating parameterized test data generators.
Unlike these approaches, our goal with \proze is not to compress or diversify the test suite.
We aim to discover CUT oracles that are well-suited for PUTs, thereby leveraging the substantial benefits that PUTs offer over CUTs.

Theory-based \cite{saff2008theory} and property-based \cite{maciver2019hypothesis, hatfield2020falsify} testing are closely related concepts to PUTs \cite{vikram2023can}.
A theory or a property is a general oracle that holds for any input data.
A key practical challenge developers face with these kinds of tests is generating realistic input data similar to the ones they would observe in the real world \cite{goldstein2024property}.
In contrast, PUTs oracles do not apply universally to all possible inputs.
They are specific to the selected set of realistic inputs provided by the argument provider.
Moreover, \proze generates argument providers that include data which both satisfy and falsify the oracle.

A key novelty of our work is our focus on realistic data captured from test and field executions.
Several studies employ runtime data for CUT generation 
\cite{wang2017behavioral, tiwari2022Pankti, tiwari2022mimicking, abdi2022dynamic}, in order to increase the input space coverage and representativeness of the test inputs.
Yet, there is no literature on using runtime data in the context of PUTs.

\section{Conclusion}
We introduce \proze, a novel technique to automatically transform CUTs into PUTs, leveraging original developer-written oracles.
To our knowledge, this is the first study on the automated generation of PUTs using realistic data observed during both test and field executions.
Our evaluation of \proze on five Java modules from two widely-used open-source Java projects demonstrates that \proze successfully handles real-world tests.
\proze captures 13,021 runtime arguments for \nbtargets target methods during the test and field executions of these five applications.
Consequently, \proze generates \nbputs PUTs from the \nbcuts CUTs that directly invoke these target methods.
Our results show that \proze significantly enhances the input space coverage for these methods.
Notably, \nbgolds PUTs are ready to be used in the projects' test suite, as they have a strong oracle that holds for a wide range of test inputs, and that can discriminate the behavior of the target method under valid and invalid inputs.

\revised{We envision that the argument providers of parameterized unit tests can include calls to faking libraries \cite{baudry2024generative}. These libraries provide reusable data for test inputs, such as names, addresses or \emph{lorem ipsum}s. We believe they can also be used to increase the coverage of input spaces, including with humorous data \cite{tiwari2024great}. In the future, we wish to assess the effectiveness of realistic runtime data and humorous fakes on the overall quality of test suites.}

\section*{Acknowledgements}
\noindent This work has been partially supported by the Wallenberg Autonomous Systems and Software Program (WASP) funded by the Knut and Alice Wallenberg Foundation, the Chains project funded by the Swedish Foundation for Strategic Research (SSF), IVADO and NSERC.

\balance
\printbibliography
\end{document}